\documentclass[prx,twocolumn,floatfix,superscriptaddress,longbibliography]{revtex4-2} 

\usepackage{mathtools}
\usepackage{amssymb,amsmath,amstext,amsthm,mathtools}
\usepackage{graphicx}
\usepackage{epstopdf}
\usepackage{color}
\usepackage{bm}
\usepackage{appendix}
\usepackage[T1]{fontenc}
\usepackage{bbold}
\usepackage{bbm}
\usepackage{latexsym}
\usepackage[colorlinks=true,citecolor=blue,linkcolor=magenta]{hyperref}
\usepackage{float}
\usepackage{verbatim}
\usepackage{lipsum}
\usepackage{braket}
\usepackage{amsthm}
\usepackage{todonotes}

\usepackage{bibentry}

\usepackage{tcolorbox}

\newcommand{\norm}[1]{\lVert #1\rVert}

\newcommand{\dya}[1]{\ket{#1}\!\bra{#1}}

\newcommand\mf[1]{\mathfrak{#1}}

\newcommand{\BC}{\mathcal{B}}

\newcommand{\EC}{\mathcal{E}}

\newcommand{\HC}{\mathcal{H}}

\newcommand{\MC}{\mathcal{M}}
\newcommand{\NC}{\mathcal{N}}
\newcommand{\OC}{\mathcal{O}}
\newcommand{\PC}{\mathcal{P}}

\newcommand{\SC}{\mathcal{S}}

\newcommand{\UC}{\mathcal{U}}

\newcommand{\XC}{\mathcal{X}}

\newcommand{\Tr}{{\rm Tr}}
\newcommand{\Var}{{\rm Var}}

\renewcommand{\geq}{\geqslant}
\renewcommand{\leq}{\leqslant}

\DeclareMathOperator*{\argmin}{arg\,min}
\renewcommand{\vec}[1]{\boldsymbol{#1}}  

\newcommand{\ad}{^\dagger}

\newcommand*{\id}{\openone}

\newcommand{\poly}{\operatorname{poly}}

\newcommand{\thv}{\vec{\theta}}

{}
{}

\theoremstyle{definition}

\definecolor{fundamental}{RGB}{55, 110, 111}

\newtcolorbox[auto counter]{pabox}[2][]{fonttitle=\bfseries,
title=Box~\thetcbcounter: #2,#1,colframe=gray}
\newtcolorbox[use counter from=pabox]{mybox}[2][]{
floatplacement=t,float,
colback=fundamental!5!white,colframe=fundamental!75!black,title=Box~\thetcbcounter: #2,#1}

\usepackage[normalem]{ulem}

\usepackage{cancel}

\begin{document}

\title{A Review of Barren Plateaus in Variational Quantum Computing}

\author{Mart\'{i}n Larocca}
\affiliation{Theoretical Division, Los Alamos National Laboratory, Los Alamos, New Mexico 87545, USA}
\affiliation{Center for Non-Linear Studies, Los Alamos National Laboratory, 87545 NM, USA}

\author{Supanut Thanasilp}
\affiliation{Institute of Physics, Ecole Polytechnique Fédérale de Lausanne (EPFL),  Lausanne CH-1015, Switzerland}
\affiliation{Chula Intelligent and Complex Systems, Department of Physics, Faculty of Science, Chulalongkorn University, Bangkok  10330, Thailand}

\author{Samson Wang}
\affiliation{Institute for Quantum Information and Matter, Caltech, Pasadena, CA 91125, USA}

\author{Kunal Sharma}
\affiliation{IBM Quantum, IBM T.J.~Watson Research Center, Yorktown Heights, NY 10598, USA}

\author{Jacob Biamonte} 
\affiliation{Quantum Artificial Intelligence Laboratory, NASA Ames Research Center, Moffett Federal Airfield, Mountain View, California 94035, USA
}
\affiliation{Simons Institute for the Theory of Computing, University of California, Berkeley, California 94720-2190, USA
}

\author{Patrick J.~Coles}
\affiliation{Normal Computing Corporation, New York, New York, USA}

\author{Lukasz Cincio}
\affiliation{Theoretical Division, Los Alamos National Laboratory, Los Alamos, New Mexico 87545, USA}
\affiliation{Quantum Science Center, Oak Ridge, TN 37931, USA}

\author{Jarrod R. McClean}
\affiliation{Google Quantum AI, Venice, CA 90291, USA}

\author{Zo\"e Holmes}
\affiliation{Institute of Physics, Ecole Polytechnique Fédérale de Lausanne (EPFL),  Lausanne CH-1015, Switzerland}

\author{M.~Cerezo}
\affiliation{Information Sciences, Los Alamos National Laboratory, 87545 NM, USA}
\affiliation{Quantum Science Center, Oak Ridge, TN 37931, USA}

\begin{abstract}
Variational quantum computing offers a flexible  computational paradigm with applications in diverse areas. However, a key obstacle to realizing their potential is the Barren Plateau (BP) phenomenon. When a model exhibits a BP, its parameter optimization landscape becomes exponentially flat and featureless as the problem size increases. Importantly, all the moving pieces of an algorithm---choices of ansatz, initial state, observable, loss function and hardware noise---can lead to BPs when ill-suited. Due to the significant impact of BPs on trainability, researchers have dedicated considerable effort to develop theoretical and heuristic methods to understand and mitigate their effects. As a result, the study of BPs has become a thriving area of research, influencing and cross-fertilizing other fields such as quantum optimal control, tensor networks, and learning theory. This article provides a comprehensive review of the current understanding of the BP phenomenon.
\end{abstract}
\maketitle

\section{Introduction}

Variational quantum algorithms have gathered momentum in recent years as a promising computational approach~\cite{cerezo2020variationalreview,bharti2021noisy,endo2021hybrid,schuld2015introduction,biamonte2017quantum,cerezo2022challenges,di2023quantum}.  At their core, most variational quantum algorithms follow
the same principle: Convert a problem into an optimization task, and use a classical device to train a parametrized quantum learning
model.  The popularity of this computational paradigm can be attributed in part to its flexibility, opening up applications in diverse areas in basic science and machine learning, but also because of the hope that it is more robust to the constraints of near-term quantum hardware.

In contrast to conventional quantum algorithms, which generally come with provable  guarantees, variational quantum algorithms share much of the heuristic nature of classical machine learning. Unlike classical machine learning, quantum computing lacks the hardware needed to test these algorithms at scale. Moreover, it quickly becomes clear to any practitioner that out-of-the-box variational schemes can be hard to train as the number of qubits becomes larger and larger.  
This has prompted efforts to analyze the scalability of these algorithms and to determine trainability barriers that could obstruct their practical use in realistic problem sizes.

Among the  sources for potential untrainability in variational quantum algorithms, this review focuses  on the \textit{Barren Plateau} (BP) phenomenon~\cite{mcclean2018barren,qi2023barren}. On a BP, the loss gradients, or more generally loss differences~\cite{arrasmith2021equivalence}, vanish exponentially with the size of the system. Combining this issue with the fact that information can only be extracted from a quantum computer through a finite set of measurements, the presence of a BP landscape implies that for the overwhelming majority of parameter choices an exponentially large number of measurement shots are needed to identify the loss minimizing direction.

By now, researchers have uncovered a number of causes of BPs, as well as several methods to mitigate or  avoid them. The quest to understand this phenomenon has highlighted that BPs are a symptom of a deeper issue with variational quantum computing. Namely, that the exponentially large Hilbert space dimension---the very same feature which was hoped to provide variational quantum computing with computational capabilities beyond those achievable by classical learning models---can lead to issues if not handled properly. That is, in essence, BPs ultimately arise from a \textit{curse of dimensionality}~\cite{cerezo2023does}. 

In this review we survey the substantial body of literature on BPs that has emerged in the last five years. Our main goal is to provide guidelines for the numerous do's and don't's that our community has identified. However, we also aim at showing that the quest to uncover the mysteries of BPs is a path filled with insights into the very nature of the information processing capabilities of quantum computers, the resources that make something quantum ``\textit{quantum}'', the connection between absence of BPs and classical simulability,  and the potential paths for alternative variational paradigms.

We start our review  in Section~\ref{sec:variational} with a brief review of variational quantum computing. In Section~\ref{sec:types-of-BPs} we recall definitions for different type of BPs, with their sources being identified in  Sections~\ref{sec:origins}. We then survey architectures (Section~\ref{sec:architectures}) and methods that can (Section~\ref{sec:do-work}) and cannot (Section~\ref{sec:not-work}) mitigate BPs, or even attempt to prevent BPs altogether. Section~\ref{sec:impact} presents examples of the impact that the BP study has had beyond the field of variational quantum computing, and Section~\ref{sec:beyond} discusses other issues beyond BPs such as local minima and the intriguing connection between absence of BPs and classical simulability. In Section~\ref{sec:classical} we link the BP phenomenon with vanishing gradients in classical machine learning. We finish in Section~\ref{sec:outlook} with some closing remarks and outlook into the future. We note that  we have added all throughout our reviews boxes with include important details and more technical discussions.

\section{Variational Quantum Computing}\label{sec:variational}

\begin{figure}[t]
    \centering
\includegraphics[width=1\linewidth]{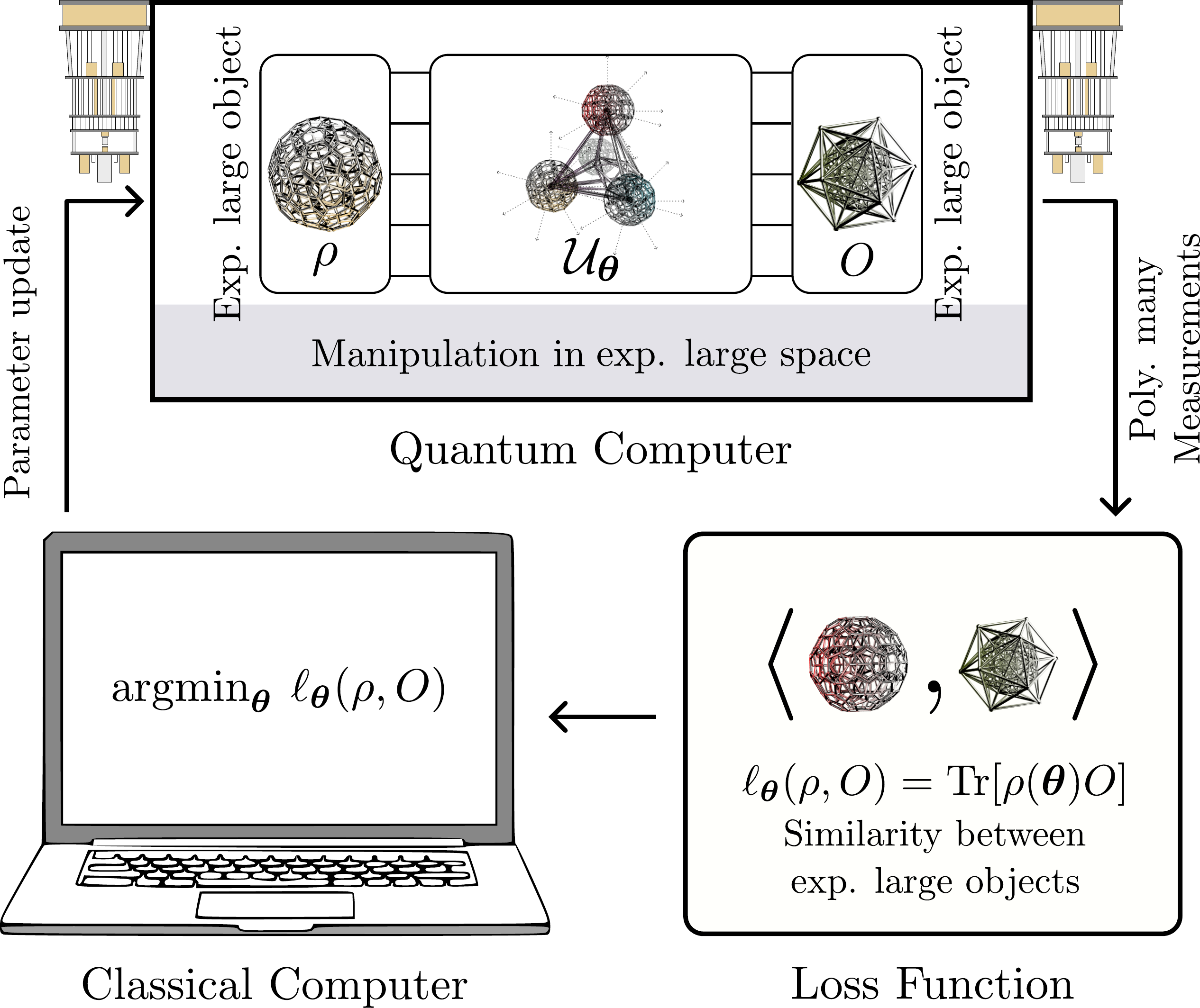}
    \caption{\textbf{Variational quantum computing.} An $n$-qubit quantum computer  is initialized to a state $\rho$ that is evolved through a Parametrized Quantum Circuit (PQC) $\UC_{\thv}$. At the output of the circuit we perform a finite set of measurements which are used to estimate a loss function $\ell_{\thv}(\rho,O)=\Tr[\UC_{\thv}(\rho)O]$. A classical computer takes as input the finite sample estimate of $\ell_{\thv}(\rho,O)$ (or its partial derivatives) and attempts to find new sets of parameters which train the PQC and minimize the loss. As such by writing the loss in the vectorized form $\ell_{\thv}(\rho,O) = \langle \rho(\thv),O\rangle$ one can see that at the core of variational quantum computing lies the manipulation and comparison---via polynomially many samples---of the exponentially large operators $\rho(\thv)=\UC_{\thv}(\rho)$ and $O$.}
    \label{fig:fig-1}
\end{figure}

\subsection{Framework}

The key idea behind variational quantum computing is to combine the  power of both quantum and classical computers.  This hybrid paradigm leverages the quantum device’s large Hilbert space to process information and estimate quantities of interest. The results are then fed to a classical computer, which analyzes them and determines the next quantum experiment. Many such hybrid algorithms employ an optimization or learning-based approach, where a classical optimizer trains quantum experiments characterized by a sequence of parameters to solve a task. Since variational computing reduces the burden on quantum hardware, it is naturally well-suited for near-term devices. However, their usage is not confined to the near-term, and variational schemes are expected to flourish in the fault-tolerant era when noise is no longer an issue.

\begin{mybox}[label={box:initial-state}]{Quantum and classical data for QML}  
By quantum  data, we mean that $\rho$ is obtained from some quantum mechanical process of interest within a  quantum device. By classical data $\vec{x}\in\XC$ we mean data from some  domain $\XC$ that can be efficiently stored in a classical computer (e.g., images, financial information, events from a particle accelerator), and that we want to process in a quantum computer. Such classical data needs to be encoded in a quantum state through an \textit{embedding map} $\EC:\XC\rightarrow \BC(\HC_0)$. We also note that the classical data $\vec{x}$ can also be used to construct the \textit{Parametrized Quantum Circuit} (PQC) in Eq.~\eqref{eq:PQC} through  data re-uploading methods~\cite{perez2020data}.
\end{mybox}

 It has become standard to divide variational quantum computing schemes into two main categories: \textit{Variational Quantum Algorithms} (VQAs)~\cite{cerezo2020variationalreview,bharti2021noisy,endo2021hybrid}  and \textit{Quantum Machine Learning} (QML) models~\cite{schuld2015introduction,biamonte2017quantum,cerezo2022challenges}. VQAs can be considered as problem-driven, since the algorithms tackle a specific task such as preparing the ground state of some Hamiltonian. QML models on the other hand are data-driven. QML has been envisioned for virtually all learning tasks such as supervised and  unsupervised learning, generative modeling, etc~\cite{cerezo2022challenges}. 
Throughout this work, we will study VQAs and QML models together, as they mostly share a common structure.

\subsection{Components of variational quantum computing: data, ansatz, measurements, and classical information processing}

As shown in Fig.~\ref{fig:fig-1}, in variational quantum computing one first initializes the quantum system to some $n$-qubit initial state $\rho$, belonging to $\BC(\HC_0)$, the set of bounded linear operators on an $2^n$-dimensional complex Hilbert space $\HC_{0}$. In VQAs, $\rho$  is usually an easy-to-prepare fiducial state, while in QML $\rho$ is taken from a set (or distribution) of training states that  encode either  \textit{quantum} or \textit{classical} data (see Box~\ref{box:initial-state}).

Next,  one sends  $\rho$ through a \textit{Parametrized Quantum Circuit} (PQC), also known as a \textit{quantum neural network} in the context of QML. A PQC can be expressed as a sequence of $L$ parametrized unitaries 
\begin{equation}\label{eq:PQC}
U(\vec{\theta})= \prod_{i=1}^L U_i(\vec{\theta}_i)\,,
\end{equation}
where $\thv=(\thv_1,\ldots,\thv_L)$ is a set of trainable parameters. When hardware noise is present, or when the parameterized quantum circuit involves adding or removing qubits, the PQC is expressed more generally as a sequence of $L$ parametrized channels
\begin{equation}
\UC_{\vec{\theta}}=\UC_{\vec{\theta}_L}\circ\cdots\circ\UC_{\vec{\theta}_1}\,,
\end{equation}
where each $\UC_{\vec{\theta}_l}:\BC(\HC_{l-1})\rightarrow \BC(\HC_{l})$ is a completely positive and trace-preserving map between bounded operators. 
Finally, as indicated in Fig.~\ref{fig:fig-1}, at the output of the PQC we perform a (finite) number of measurements to estimate a quantity of interest such as the expectation value of some set of observables or the probability of a given outcome.

\begin{mybox}[label={box:inductive-biases}]{Inductive biases} 
Inductive biases, broadly speaking, capture the assumptions and choices we make when constructing the quantum algorithm. In the pipeline of variational quantum computing these come in many forms. For instance, when working with VQAs, while there is a significant amount of freedom when choosing the input state,  the measurement operator is usually more restricted. This is due to the fact that in most VQAs the problem information is encoded in the measurement operator (e.g., in the variational quantum eigensolver~\cite{peruzzo2014variational} we measure the expectation value of the Hermitian operator whose ground state we want to prepare). In QML, the input states tend to be fixed, with the encoding scheme for classical data, the measurement operator, and the choice of loss function being generally more flexible, offering additional design freedom."Similarly, while Eq.~\eqref{eq:PQC} shows a general  PQC, different instantiations of these computational models differ based on the inductive biases that influence $\UC_{\vec{\theta}_l}$, i.e., the choices in their specific form. This includes properties such as the number and types of gates, their placement, and how the parameters are initialized. 
\end{mybox}

In the simplest case, one has a single input state and is interested in estimating the expectation value of a single observable $O$, which leads to the loss function 
\begin{equation}\label{eq:loss-funcion}
    \ell_{\thv}(\rho,O)=\Tr[\rho(\thv)O]\,,
\end{equation}
with $\rho(\thv)=\UC_{\vec{\theta}}(\rho)$. While more general loss functions will be discussed below (some built by computing several quantities such as $\ell_{\thv}(\rho,O)$), we will mostly  restrict our attention to those of the form in Eq.~\eqref{eq:loss-funcion}, as the lessons learnt therein can often be extrapolated to other, more general, scenarios. In either case, one usually uses a quantum device to estimate the loss, and leverages the power of classical optimizers to solve the optimization task 
\begin{equation}\label{eq:opt-task}
    \argmin_{\thv} \ell_{\thv}(\rho,O)\,.
\end{equation} 
Despite the (relative) simplicity of the variational quantum computing framework, solving Eq.~\eqref{eq:opt-task} and training the model's parameters can be a hard task. For instance, the optimizer might get stuck in a sub-optimal local minima~\cite{bittel2021training,fontana2022nontrivial,anschuetz2022beyond,anschuetz2021critical,larocca2021diagnosing,bermejo2024improving,nadori2024promising}, there might not exist a solution in the parameter hyperspace~\cite{sim2019expressibility,akshay2020reachability}, or the landscape might exhibit BPs, the focus of this review. Importantly, and as we will see below, the hardness of training the loss will strongly depend on the inductive biases of the quantum algorithm (see Box~\ref{box:inductive-biases}).

\begin{mybox}[label={box:finite-sampling}]{Effects of finite sampling} 
In practice, we can only estimate the expectation value of the loss function $\ell_{\thv}(\rho,O)=\mathbb{E}_{\rho(\thv)}[O]$ up to some finite precision determined by the number of measurement shots. To differentiate, $\ell_{\thv}(\rho,O)$ from its $N$-shot approximation, we  use $\overline{\ell}_{\thv}(\rho,O)$ to denote the latter. Assuming that the experiment repetition used to obtain $\overline{\ell}_{\thv}(\rho,O)$ are independent, we recover the sample variance (denoted as $\Var_s$, and also known as the standard deviation of the mean~\cite{samplevariance}) of the estimate 
\begin{equation}\label{eq:sample-variance}
\begin{split}
    \Var_s[\overline{\ell}_{\thv}(\rho,O)]&=\frac{\Var_{\rho(\thv)}[O]}{N}\\
&=\frac{\mathbb{E}_{\rho(\thv)}[O^2]-\mathbb{E}_{\rho(\thv)}[O]^2}{N}\,.
\end{split}
\end{equation}
Equation~\eqref{eq:sample-variance}  quantifies the uncertainty to which $\overline{\ell}_{\thv}(\rho,O)$ is computed. Clearly, there are two contributions to this uncertainty. First, the one in the numerator arises from the fact that $\rho(\thv)$ is generally not an eigenstate of $O$, and hence different measurements can yield different outcomes. Second, the factor $1/N$ indicates that we are estimating the expectation value with a finite number of samples, and therefore. Via Chebychev's inequality this implies  that the statistical fluctuations of the mean estimator are of order $1/\sqrt{N}$.  
Thus, in regions of a landscape with exponentially small gradients, exponentially many shots are typically required to determine a loss minimizing direction. 
\end{mybox}

\section{Types of Barren Plateaus and Loss Concentration}\label{sec:types-of-BPs}

We define a barren plateau (BP) in a variational quantum computing scheme as occurring when the loss function, or its gradients, become exponentially concentrated around their mean as the number of qubits $n$ increases. Intuitively, this means that the optimization landscape is mostly flat and featureless and that slightly changing the model's parameters $\thv$ results in only an exponentially small change in $\ell_{\thv}(\rho,O)$ or $\partial_\mu\ell_{\thv}(\rho,O)=\partial\ell_{\thv}(\rho,O)/\partial\theta_\mu$ (where $\theta_\mu\in\thv$). This creates significant challenges for minimizing the loss function, as optimizers determine descent directions by comparing the loss at different points. In particular, since  we can only estimate expectation values from a quantum computer with a finite number of measurements $N$, these come with a statistical uncertainty that scales inversely proportional to $\sqrt{N}$ (see Box~\ref{box:finite-sampling}).  To resolve exponentially small changes and optimize the loss, we would need an exponentially large number of measurements (shots), which makes the algorithm inefficient and non-scalable. Indeed, if a smaller number of shots are used, the optimizer will  follow changes produced by the uninformative statistical fluctuations, meaning that the parameter updates lead to a useless random walk in the loss landscape.

\subsection{Probabilistic concentration and narrow gorges}

The most common type of loss function concentration encountered in the literature is known as \textit{probabilistic} concentration. 
By defining $\mathbb{E}_{\thv}$ as the expectation value with respect to $\thv$ sampled from some domain according to some distribution, we say a loss exhibits a probabilistic BP if
\begin{equation}\label{eq:concentration-var}
\Var_{\thv}[\ell_{\thv}(\rho,O)]\; \text{or} \; \Var_{\thv}[\partial_\mu\ell_{\thv}(\rho,O)]\in\OC\left(\frac{1}{b^n}\right)\,,
\end{equation}
for some $b>1$, and for some but (not necessarily all)  parameters $\theta_\mu \in \vec{\theta}$. 
As discussed later in Sec.~\ref{sec:origins}, the choice of an ansatze $U(\thv)$, an initial state $\rho$ and an observable $O$ can individually induce the concentration. The exact value of the exponent base $b$ implicitly depends on $U(\thv)$, $\rho$ and $O$.

At this point, we note that one should understand probabilistic BPs as an \textit{average} properties of a loss landscape, which means that ``most'' points in the landscape should be exponentially flat. To see this more formally, the exponential vanishing variance implies through Chebyshev's inequality that 
for any $\delta>0$ 
\begin{equation}\label{eq:probab-loss}
{\rm Pr}_{\thv}(|\ell_{\thv}(\rho,O)-\mathbb{E}_{\thv}[\ell_{\thv}(\rho,O)]|\geq \delta)\in \OC\left(\frac{1}{b^n}\right)\,,
\end{equation}
or ${\rm Pr}(|\partial_\mu\ell_{\thv}(\rho,O)-\mathbb{E}_{\thv}[\partial_\mu\ell_{\thv}(\rho,O)]|\geq \delta)\in \OC\left(\frac{1}{b^n}\right)$. 
These equations show that the probability of the loss or its partial derivative deviating by more than $\delta$ from its mean is exponentially small, where $\delta\in\Omega(1/\poly(n))$. 

Here we find it important to remark that while the definition of a BP in terms of the loss is clear, it becomes less clear when expressed in terms of partial derivatives. As mentioned, a BP arises when some, but not all, partial derivatives are concentrated, indicating that analyzing the concentration of the overall loss is a more reliable way to diagnose a BP.  In fact, it is entirely possible that a loss  contains some partial derivatives that are concentrated while others are not. Hence,  care must be taken when establishing the absence of a BP by studying $\Var_{\thv}[\ell_{\thv}(\rho,O)]$ as it is entirely possible that a non-concentrated loss arises only from a few well-behaved set of parameters (and thus still has a BP).  
However, when \textit{every} partial derivative is concentrated one can prove that loss concentration and partial derivative concentration are equivalent~\cite{arrasmith2021equivalence,miao2024equivalence}. Similarly, one can show that if first-order derivatives concentrate, then so do higher-order ones~\cite{cerezo2020impact}.

Although a probabilistic BP implies a largely featureless landscape, regions with large gradients around some minima (sometimes referred to as fertile valleys~\cite{valls2024variational}) can still exist. The caveat is that these regions have to be small and the  minima therein are always exponentially narrow (meaning that their \textit{relative} volume in parameter space is always exponentially small~\cite{arrasmith2021equivalence}). These properties led researchers to name these minima as \textit{narrow gorges}. While this term suggests a steep well, the slope of the walls in these gorges is bounded by the loss’s Lipschitz constant, which is typically in  $\OC(\poly(n))$~\cite{arrasmith2021equivalence}.  Lastly, since these are the most common BPs discussed in this review, we henceforth assume BPs to be probabilistic unless stated otherwise.

\subsection{Deterministic concentration}

While probabilistic BPs lead to  mostly flat and featureless landscapes, they still admit the presence of well-behaved regions such as narrow gorges. In contrast, there exists a second type of concentration, known as \textit{deterministic} BPs, where all of the landscape is truly flat~\cite{wang2020noise,diaz2023showcasing,leone2022practical,thanasilp2021subtleties}. A deterministic BP arises when the loss function’s deviation from its mean is bounded for all parameter values, $\forall \theta$, by an exponentially small term. That is, when
\begin{equation}\label{eq:deterministic-loss}
|\ell_{\thv}(\rho,O)-\ell_0]|\in\OC\left(\frac{1}{b^n}\right)\,,
\end{equation}
for some $b>1$, and where $\ell_0$ is a concentration point  which is usually independent of $\thv$.  Deterministic concentration thus arises due to the state $\rho(\thv)$ only having exponentially small overlap with $O$ for all values of $\thv$ (which therefore implies that this BP phenomenon depends on the initial state and measurement operator).  As further discussed below, this situation can arise due to 
high entanglement in an input state (see Sec.~\ref{sec:input-state-measurement}) and/or unital noise in PQC (see Sec.~\ref{sec:noise}). Clearly, Eq.~\eqref{eq:deterministic-loss} implies that for all parameter values $|\partial_\mu\ell_{\thv}(\rho,O)-\ell_0|\in\OC\left(\frac{1}{b^n}\right)$. A deterministically concentrated loss cannot have any extrema that are significantly separated from the loss’s mean. In fact, Eq.~\eqref{eq:deterministic-loss} precludes any type of non-exponentially suppressed features in the landscape.

\subsection{Methods for analyzing variance}

At the simplest level, one can determine if the loss or its gradients are concentrating by evaluating randomly sampled parameters and computing how the variance scales with the number of qubits~\cite{mcclean2018barren,cybulski2023impact,larocca2021diagnosing}. Similarly, one can also compute other quantities proposed as surrogates to study whether a BP occurs These include  the landscape's information content~\cite{perezsalinas2023analyzing} and fluctuation~\cite{zhang2024predicting}, the local state's entanglement~\cite{sack2022avoiding}, or the landscape's Fourier coefficients~\cite{okumura2023fourier,nemkov2023fourier,stkechly2023connecting}. While these approaches can provide a rough estimate of scaling, they suffer from statistical errors due to finite measurements of the loss and generally cannot distinguish between the two types of BPs.

A preferred, but more theoretically demanding, method for detecting BPs is by  analytically studying the scaling of the variance. In this method, assumptions about the PQC are necessary to make the problem more tractable. For instance, if  $\mathcal{U}_{\boldsymbol{\theta}}$ is a deep unitary circuit whose distribution of unitaries forms at least a $2$-design~\cite{dankert2009exact} over a Lie group (for a brief introduction to the topic, see Box~\ref{box:Lie} for Lie algebras, Box~\ref{box:groups} for an average over the group, and Box~\ref{box:expressiveness} for the PQC expressivity), one can use representation theory to exactly evaluate the variances~\cite{mcclean2018barren,sharma2020trainability,patti2020entanglement,marrero2020entanglement,holmes2021barren,larocca2021diagnosing,ragone2023unified,fontana2023theadjoint,diaz2023showcasing,schatzki2022theoretical,liu2022laziness,west2023provably,garcia2023deep} via Weingarten calculus~\cite{mele2023introduction,ragone2022representation}.  Instead, for shallow circuits that do not form designs, one can instead assume that the local gates are sampled from a local group, thus mapping the variance evaluation to a Markov chain-like process which can be analyzed analytically~\cite{cerezo2020cost,uvarov2020barren,pesah2020absence,heyraud2023efficient,liu2021presence,barthel2023absence,miao2023isometric,garcia2023barren,letcher2023tight,arrazola2022universal,zhang2023absence,garcia2024architectures,deneris2024exact}, numerically via Monte Carlo~\cite{napp2022quantifying} and tensor networks~\cite{braccia2024computing,hu2024demonstration}, or studied via XZ-calculus~\cite{zhao2021analyzing}. One can also take an iterative approach to explicitly integrate over the random parameters  within a given angle range~\cite{zhang2022escaping, wang2023trainability, letcher2023tight, valls2024variational} or exploit tools from Floquet theory~\cite{park2023hamiltonian, park2024hardware}. In addition, if $\mathcal{U}_{\boldsymbol{\theta}}$ is a noisy channel, one can study the variance by making assumptions regarding to the type of noise acting throughout the PQC~\cite{wang2020noise,sannia2023engineered,liu2022noise,schumann2023emergence,singkanipa2024beyond}.

\begin{mybox}[label={box:Lie}]{From quantum circuits to Lie algebras and groups}

The Lie subalgebras $\mf{g}\subseteq \mf{su}(\HC)$ play a fundamental role in BP analysis.
Recalling that $\mf{su}(\HC)$ is the subspace of traceless anti-Hermitian linear operators on $\HC$, any subspace $\mf{g}\subseteq\mf{su}(\HC)$ that is closed under commutation constitutes a subalgebra. Given a unitary noiseless PQC as in Eq.~\eqref{eq:PQC} where $U(\thv_l) = e^{-i \theta_l H_l}$, its \textit{dynamical Lie algebra} (DLA)~\cite{zeier2011symmetry,larocca2021diagnosing,wiersema2023classification,wierichs2023symmetric} is defined as
\begin{equation}
    \mf{g} = \langle \{iH_l\}_l \rangle_{{\rm Lie}} \subseteq \mf{su}(\HC)\,,
\end{equation}
where $\langle S \rangle_{{\rm Lie}}$, for any subset $S\subseteq \mf{su}(\HC)$, denotes its \textit{Lie closure}, the real vector space spanned by every nested commutator between the elements of $S$. The DLA is a fundamental object in the study of PQCs because its associated Lie group $G=e^{\mathfrak{g}}$ contains all unitaries that can ever be achieved by a choice of parameters $\thv$ or number of layers $L$. One can explicitly verify this by using the Baker–Campbell–Hausdorff formula to re-express the composition of layers
as the exponential of a linear combination of the nested commutators of the generators, which by definition belongs in $\mf{g}$.
\end{mybox}

\section{Origins of BPs}\label{sec:origins}

Theoretical and numerical calculations can help determine the flatness of the landscape, but they merely reveal symptoms of deeper issues in the variational quantum computing scheme. In this section we discuss recent findings which argue that BPs appear as a form of curse of dimensionality, as well as showcase how this phenomenon reveals itself under different scenarios. By understanding the connections between these different origins of BPs, we will see that one should be able to predict if a model will exhibit an exponentially concentrated loss.

\begin{mybox}[label={box:groups}]{Averages $\&$ Haar-measure $\&$  Weingarten calculus}
Evaluating $\mathbb{E}_{\thv}[\Tr[U(\thv)\rho U\ad(\thv)O]^t]$ can become  tractable for unitary PQCs under certain assumptions. First, note that every set of parameters $\thv$, determines a unitary through Eq.~\eqref{eq:PQC}, meaning that we can opt to evaluate the average over the induced  distribution of unitaries $\SC$, and an associated distribution $dU$ over $\SC$. That is, we can evaluate $\mathbb{E}_{\thv}[\Tr[U(\thv)\rho U\ad(\thv)O]^t]$ as  
\begin{align}\label{eq:twirl}
    \int_{\SC}dU[\Tr[U\rho U\ad O]^t]&=\Tr[\tau_{\SC}^{(t)}(\rho^{\otimes t})O^{\otimes t}]\,,
\end{align}
where $\tau_{\SC}^{(t)}(\cdot)= \int_{\SC}dUU^{\otimes t}(\cdot)(U^*)^{\otimes t}$ is the $t$-th fold twirl over $\SC$. (In the vectorized picture~\cite{mele2023introduction}, we can also write Eq.~\eqref{eq:twirl} as $\langle\langle \rho^{\otimes t} |\widehat{\tau}_{\SC}^{(t)}|O^{\otimes t}\rangle\rangle$ with $\widehat{\tau}_{\SC}^{(t)}=\int_{\SC}dUU^{\otimes t}\otimes (U^*)^{\otimes t}$  known as the $t$-th moment operator.)  If $\SC$ forms a group $G$, and if $dU$ is the uniform measure over this group, we say that the PQC is Haar random over $G$, in which case $\tau_{\SC}^{(t)}$ (or $\widehat{\tau}_{\SC}^{(t)}$) can be evaluated analytically through the Weingarten calculus. Specifically, given a basis $\{B_\sigma\}_\sigma$ for the $t$-th fold commutant of $G$,  ${\rm comm}^{(t)}(G)=\{A\in\BC(\HC^{\otimes t})\,|\, [A,U^{\otimes t}]=0\,\,\forall U\in G\}$, we have 
\begin{equation}
    \tau_{\SC}^{(t)}(\cdot)=\sum_{\sigma \pi}W^{-1}_{\pi\sigma }\Tr[B_{\sigma}\ad (\cdot)]B_{\pi}\,,
\end{equation}
where $W$ is the basis' Gram matrix. For a pedagogical introduction to Weingarten calculus with a focus in quantum information we point the reader to~\cite{mele2023introduction}.
\end{mybox}

\subsection{A curse of dimensionality}\label{sec:curse-of-dim}

Recently it has been argued that the very same feature that makes variational quantum computing appealing---the exponentially large dimension of the Hilbert space---is at the very core of the BP phenomenon~\cite{ragone2023unified,fontana2023theadjoint,diaz2023showcasing,cerezo2023does,bremner2009random,gross2009most}.   To see that this is the case, let us begin by assuming that the PQC is  a unitary as in Eq.~\eqref{eq:PQC}. Then, a loss of the form in Eq.~\eqref{eq:loss-funcion} can be expressed as $\ell_{\thv}(\rho,O)=\langle \rho(\thv),O\rangle=\langle \rho,O(\thv)\rangle$, where we introduced $O(\thv)=U(\thv)\ad O U(\thv)$ and with $\langle A,B\rangle=\Tr[A\ad B]$. That is, the loss is the Hilbert-Schmidt inner product between two vectors in the exponentially large space $\BC(\HC_0)$. This simple  realization shows that  optimizing the loss can be understood as variationally trying to anti-align two exponentially large vectors via their inner product.  This is already troublesome as---under quite general assumptions such as random initialization---the inner product between two exponentially large parametrized objects will be (on average) exponentially small and concentrated.

Now, of course, this does \textit{not} mean that comparing \textit{any} pair of exponentially large objects will always lead to exponentially small signals. Indeed, many classical machine learning algorithms~\cite{yang2022tensor}, as well as conventional quantum algorithms~\cite{nielsen2000quantum} manipulate vectors in the exponentially large Hilbert space by carefully choreographing the patterns of constructive and destructive interference of the probability amplitudes~\cite{horgan2016scott}. This is often because these methods incorporate specific structures or constraints that limit their expressiveness or
control the exploration of the parameter space, thereby avoiding uncontrolled concentration effects. Moreover, it is also worth noting that conventional quantum algorithms have even been adapted to establish advantage proofs for quantum machine learning~\cite{liu2021rigorous, jager2023universal}. 
Yet, there are not many such algorithms in the variational setting as their construction requires carefully and purposely designed circuits with appropriate inductive biases for the task they are trying to solve. How well variational quantum algorithms can be coerced into finding and exploiting such structure in the exponentially large Hilbert space is very much an open question. 

In what follows, we will take a closer look at $\langle \rho(\thv),O\rangle$ and illustrate how its different components can lead to a curse of dimensionality-induced exponential concentration due to the choice of circuits, initial states, measurement operators, and also due to the effects of hardware noise. 

\subsection{Circuit expressiveness}

\begin{mybox}[label={box:expressiveness}]{Expressive power of the PQC} 
The expressive power of a unitary PQC quantifies the breath of unitaries that the circuit is likely to produce in an unbiased manner. For instance, one can quantify the \textit{potential}, or \textit{ultimate} expressiveness of the PQC by its dynamical Lie algebra $\mathfrak{g}$ defined in Box~\ref{box:Lie}---since  any unitary $U(\thv)$ produced by the PQC belongs in the dynamical Lie group $G=e^{\mathfrak{g}}$. Here, the larger the dimension of the algebra, the more expressive the circuit is, with controllable circuits being defined as those for which $\mathfrak{g}=\mathfrak{su}(2^n)$. Of course, shallow circuits will not be able to cover the whole dynamical Lie group $G$, as they can only produce unitaries in some subset $\SC\subseteq G$. To deal with this situation, we can instead consider the \textit{current} expressiveness which is quantified by the difference between $\tau_{\SC}^{(t)}$ and $\tau_{G}^{(t)}$ (where we recall that these quantities are defined in Box~\ref{box:groups}). When those operators match, we say that the PQC forms a $t$-design~\cite{dankert2009exact}. Importantly,  if $|\tau_{\SC}^{(2)}-\tau_{G}^{(2)}|\leq \varepsilon$ (in operator norm) then~\cite{ragone2023unified}
\begin{equation}
    |\Var_{\thv}[\ell_{\thv}(\rho,O)]-\Var_{G}[\ell_{\thv}(\rho,O)]|\in\OC( \varepsilon\norm{O}_1)\,, \nonumber
\end{equation}
showing that we can quantify how much the variance of a shallow circuit deviates from that of a Haar random PQC over $G$ via  $\varepsilon$.
\end{mybox}

Perhaps the main culprit responsible for the curse of dimensionality is the PQC as it has been shown that its expressive power (see Box~\ref{box:expressiveness}) is directly linked to BPs. Roughly speaking, the larger the space the
PQC explores in an unbiased manner, the more it is prone --when randomly initialized-- to lead to input
states that have exponentially small inner products with
the measurement operator. The previous connection between expressiveness and BPs has been mathematically formalized for unitary~\cite{mcclean2018barren,cerezo2020cost,holmes2021connecting,larocca2021diagnosing,ragone2023unified,fontana2023theadjoint,diaz2023showcasing,friedrich2023quantum} and noisy~\cite{duschenes2024channel} circuits, as one can show that  the more expressive a circuit is, the more the loss can concentrate. Similarly, the PQC's capacity ~\cite{abbas2020power,haug2021capacity} --measured by how many functions it can represent--affects the concentration of the loss. The higher the capacity, the more likely it is that the loss will concentrate~\cite{larocca2021theory}.

In particular, let us review the recent approach in Refs.~\cite{ragone2023unified,fontana2023theadjoint,diaz2023showcasing} for obtaining an exact analytical form for the loss function variance of deep unitary PQCs. By \textit{deep} we refer to the regime where the number of layers is large enough to guarantee that the distribution
of unitaries corresponding to random parameter choices forms a $2$-design~\cite{dankert2009exact}. That is, that it approximately matches the Haar distribution on
$e^{\mf{g}}$ (where $\mf{g}$ is the circuit's \textit{dynamical Lie algebra} (DLA), defined in Box~\ref{box:Lie}) up to second moments with small additive approximation error.  
Assume for simplicity that either $\rho$ or $O$ belong to a group-module $\MC\subseteq \BC(\HC)$ that is non-trivial under the adjoint action of the circuit. (We recall that a group-module is a vector space closed under the action of the group. In the case of $\BC$, $\MC$ is a subspace that is closed under the adjoint action of $e^{\mathfrak{g}}$, i.e., if $A\in\MC$, then $UAU\ad$ also belongs to $\MC$ for any $U\in e^{\mathfrak{g}}$~\footnote{Technically speaking, group modules are irreducible representations of the dynamical Lie group over the vector space $\BC(\HC_0)$}). One finds 
\begin{equation}\label{eq:var-DLA}
    \Var_{\thv}[\ell_{\thv}(\rho,O)]=\frac{P_{\MC}(\rho)P_{\MC}(O)}{\dim({\MC})}\,.
\end{equation}
Here, $P_{\MC}(\rho)$ and $P_{\MC}(O)$ are the norms of the projection of $\rho$ and $O$ onto the module $\MC$. Explicitly,  given an Hermitian orthonormal basis for $\MC$, denote as $\{H_\mu\}_\mu$, then $P_{\MC}(\rho)=\sum_\mu\Tr[H_\mu\rho]^2 $. As one can see, and as further discussed below,  these three quantities on the right hand side of Eq.~\eqref{eq:var-DLA} are associated with different in which  BPs can arise. In particular, expressive power of PQC is related to $\dim({\MC})$ and the role of an input state as well as measurements are associated with $P_{\MC}(\rho)$ and $P_{\MC}(O)$, respectively. Having a PQC with exponentially large $\dim({\MC})$ and/or exponentially small $P_{\MC}(\rho)$ and/or $P_{\MC}(O)$ results in an exponentially small variance and hence BPs.

To finish, let us note that in order to obtain Eq.~\eqref{eq:var-DLA} we assumed that the circuit is deep enough so that it forms a $2$-design over $e^{\mathfrak{g}}$. The number of layers $L$ for which this assumption is satisfied can be quantified via the results in~\cite{ragone2023unified,fontana2023theadjoint}. Interestingly, one can still bound how much the variance will deviate from that in Eq.~\eqref{eq:var-DLA} when the circuit is not a $2$-design (see for instance Box~\ref{box:expressiveness} and the results in~\cite{ragone2023unified,fontana2023theadjoint,holmes2021connecting,friedrich2023quantum}). Moreover, for shallow circuits where Eq.~\eqref{eq:var-DLA} does not hold, the computation of the loss function variance can become more intricate but a similar idea holds: If the circuit's adjoint action can move the measurement operator (or the initial state) in an exponentially large subspace of operator space, then a BP can arise.

\subsection{Input states and measurements}\label{sec:input-state-measurement}

While highly expressive PQCs significantly contribute to BPs, circuits with limited expressiveness can also exhibit BPs depending on the initial state and measurement operator.  For instance, let us consider Eq.~\eqref{eq:var-DLA} again, and assume that the ensuing DLA is not exponentially large, i.e., that we have $\dim(\mathfrak{g})\in\OC(\poly(n))$. This occurs when $\UC_{\thv}$ has strong inductive biases such as encoding some symmetries (see below for specific examples). Now, despite using a circuit with a non-exponentially large underlying algebra, the loss can still be exponentially concentrated. For instance, if we pick $O$ belonging to an exponentially large module (i.e., small algebras can admit exponentially large modules), then one will find exponential concentration from the denominator in Eq.~\eqref{eq:var-DLA}. However, if $O$ belongs to a module whose dimension grows polynomially with $n$, such as $iO\in\mathfrak{g}$, the variance’s denominator $\dim(\MC)=\dim(\mathfrak{g})$ will only decrease polynomially with $n$, meaning that the circuit expressiveness cannot be responsible for a BP. In this case, the loss is obtained by comparing objects in a polynomially large subspace, potentially avoiding the curse of dimensionality. However, a deterministic BP is still possible if the norm of the projection of $\rho$ into the module is exponentially small, i.e., if $\PC_{\MC}(\rho)\in\OC(1/2^n)$. Box~\ref{box:single-qubit} presents an example where an inexpressive circuit can still lead to a BP, either because $O$ belongs to an exponentially large $\MC$ or because $O$ comes from a polynomial module but the state is misaligned.

We refer the reader to Refs.~\cite{ragone2023unified,diaz2023showcasing} for a conceptual discussion on the interplay between PQC, measurement and initial state. Therein, it was argued that the module purities $\PC_{\MC}(\rho)$ can be considered as generalized forms of entanglement~\cite{somma2004nature,somma2005quantum}, thus providing operational meaning to those quantities. Moreover, Refs.~\cite{khatri2019quantum,cerezo2020cost,uvarov2020variational,uvarov2020barren,kashif2023impact,leadbeater2021f,cerezo2020variational,kiani2022learning,zambrano2023avoiding,ogunkoya2023investigating} and~\cite{marrero2020entanglement,patti2020entanglement,sack2022avoiding,zhang2023energy} studied problems where the measurement operator or the state, respectively, can determine the presence of a BP. Then, Ref.~\cite{wiersema2023measurement} showed that mid-circuit measurements can change the amount of entanglement in the state and thus determine whether a BP will arise. For the case of variational quantum computing with classical data, it has been shown that the encoding scheme can lead to initial states that are ill-aligned with the PQC and measurement operator, further highlighting the importance of choosing a well-design data embedding scheme~\cite{thanasilp2021subtleties,shaydulin2021importance,abbas2020power,schatzki2022theoretical,kashif2023unified,das2023role,mhiri2024constrained}.

\begin{mybox}[label={box:single-qubit}]{Single-qubit rotation PQCs can have BPs} 
Consider a unitary circuit composed of general single qubit gates as $U(\thv)=\prod_{j=1}^n U_j(\thv_j)$, where $U_j(\thv_j)$ acts only on the $j$-th qubit. Clearly, this circuit is fairly inexpressive as it does not even generate entanglement. Moreover, its DLA is simply given by $\mathfrak{g}=\oplus_{j=1}^n \mathfrak{su}(2)$, and thus of dimension $\dim(\mathfrak{g})=3n\in\OC(\poly(n))$.  Yet, we can see that if $O=Z^{\otimes n}$, then the operator belongs to an exponentially large module composed of all $3^n$ Pauli operators acting non trivially on all the qubits. Assuming that each $U_j(\thv_j)$ forms a $2$-design over $\mathbb{SU}(2)$~\cite{ragone2023unified}, a direct calculation reveals that for all $\rho$,  $\Var_{\thv}[\ell_{\thv}(\rho,O)]\in\OC(1/2^n)$. However, if  $O=Z_1$, then the measurement belongs to a module composed of the $3$ Pauli operators acting on the first qubit. Now, Eq.~\eqref{eq:var-DLA} leads to $\Var_{\thv}[\ell_{\thv}(\rho,O)]=2(\Tr[\rho_1^2]-1/2)/3$, with $\rho_1$ the reduced state of $\rho$ on the first qubit. The choice of measurement operator can lead to a BP. When $O=Z^{\otimes n}$ the loss is obtained by comparing operators that can move in exponentially large modules, while for $O=Z_1$, the circuit can only move in the subspace of operators acting only on the first qubit. However, while the PQC and measurements are not responsible for BPs, the initial state can lead to an exponentially concentrated loss. In particular, if the state is too entangled and $(\Tr[\rho_1^2]-1/2)\in\OC(1/2^n)$ (which occurs if $\rho$ satisfies a volume law of entanglement~\cite{leone2022practical}), then a deterministic BP occurs. 
\end{mybox}

\subsection{Noise}\label{sec:noise}

While in the previous sections we have considered unitary circuits, the presence of hardware noise throughout the PQC can significantly affect the loss function optimization landscape, potentially inducing BPs~\cite{wang2020noise,franca2020limitations,sannia2023engineered,liu2022noise,schumann2023emergence,singkanipa2024beyond,singkanipa2024beyond, depalma2022limitations, crognaletti2024estimates}. When modeling noise, it is standard to assume that the channels $\UC_{\thv_l}$ consist of a unitary parametrized gate followed by a noise channel $\NC$ as $\UC_{\thv_l}=\NC(U_l(\thv_l)\rho U_l\ad(\thv_l))$. In this picture, it has been shown that any noisy PQC  whose noise channels have the maximally mixed state as its fixed point (i.e., such that $\NC\circ\cdots\circ\NC(\rho)\sim\id/2^n$) will induce a deterministic BP in the large number of layers regime (see Box~\ref{box:NIBP} for an example with global depolarizing noise)~\cite{wang2020noise,franca2020limitations}. Recent findings show that non-unital noise or intermediate measurements can alleviate the BP issue, suggesting that the relationship between noise, measurements, and loss function concentration remains an open question~\cite{sannia2023engineered,mele2024noise,fefferman2023effect,duschenes2024channel, crognaletti2024estimates,deshpande2024dynamic}.

\begin{mybox}[label={box:NIBP}]{Noise-induced barren plateaus} 
Let us consider a noisy PQC which we model as unitary gates interleaved with global depolarizing noise channels $\NC(\rho)=(1-p)\rho+p\id/2^n$~\cite{wilde2013quantum,nielsen2000quantum}. That is, we will write $\UC_{\thv_l}(\rho)=(1-p)U(\thv_l)\rho U\ad (\thv_l) + p\id/2^n $, which indicates that with probability $(1-p)$ the unitary gate acts, whereas with probability $p$ the state is fully depolarized into the maximally mixed state. One can readily verify that the loss becomes $\ell_{\thv}(\rho,O)=(1-p)^L \Tr[ U(\thv)\rho U\ad(\thv)O]+ (1-(1-p)^L)\Tr[O]/2^n$, implying a deterministic concentration (towards $\Tr[O]/2^n$) as the circuit's depth $L$ increases or as the depolarizing probability  becomes one. Note that this noise-induced BP will appear irrespective of whether the noiseless  loss function is exponentially concentrated or not. That is, even if the noise-free loss landscape does not have BP i.e., $\Var_{\thv}\left[\Tr[ U(\thv)\rho U\ad(\thv)O]\right]\in\Omega(1/\poly(n))$, the presence of global depolarizing noise can lead to a BP if the circuit depth $L$ is sufficiently large.
\end{mybox}

While not obvious a priori, one can  see that in some cases noise-induced BPs still count as a form of curse of dimensionality~\cite{duschenes2024channel}. For instance, by recalling that noisy processes can always be understood as an entangling operation with some environment of $n_E$ qubits. Thus, if we purify~\cite{nielsen2000quantum,wilde2013quantum} the noisy state $\rho_{\thv}=\Tr_E[V_{\thv} (\dya{0}^{\otimes n_E}\otimes \rho)V\ad_{\thv}]$  we see that the loss becomes $\ell_{\thv}(\rho,O)=\langle V_{\thv} (\dya{0}^{\otimes n_E}\otimes \rho)V\ad_{\thv},\id_E\otimes O\rangle$, with $\id_E$  the identity over the environment's Hilbert space. Thus, noise leads to an extended PQC $V_{\thv}$ which can act in a larger space (that could be exponentially large, although not necessarily), and thus substantially aggravates the curse of dimensionality.

\section{Barren-plateau-prone architectures}\label{sec:architectures}

In the previous section we have provided some intuition regarding the origins of  BPs. In this section we  review some BP-prone architectures.  We note that in most cases, deep versions of these architectures can lead to exponentially concentrated losses due to their large DLAs. However, these architectures can still be useful in shallow circuits or when using smart initialization strategies.

\subsection{Hardware efficient ansatz and deep unstructured circuits}

Hardware efficient ansatz~\cite{kandala2017hardware} is a generic term  employed for unstructured PQCs composed of single-qubit  rotation gates interleaved with (fixed or parametrized) entangling gates. The term hardware-efficient spans from the fact that the gates of the circuit are chosen not because they encode some strong inductive bias from the problem one is trying to solve, but because they are easy to implement on the device (e.g., one only picks entangling gates that follow the device connectivity to avoid compiling overhead~\cite{khatri2019quantum,he2021variational,moro2021quantum}). Under quite general conditions, hardware efficient ansatzes will be universal ~\cite{larocca2021diagnosing}, meaning that their DLA is $\mathfrak{g}=\mathfrak{su}(2^n)$. As such,  deep versions of these circuits are as expressive as possible, and therefore will exhibit BPs irrespective of the initial state or measurement operator~\cite{beer2020training,sharma2020trainability,mcclean2018barren,cerezo2020cost,buonaiuto2024effects,liu2022mitigating,zhang2024exponential}. Here it is important to note that several standard architectures exist in the literature, such as entangling gates acting on a brick-like fashion on alternating pairs of qubits or on a ladder. While these can indeed exhibit different performances in their shallower forms, they will still lead to a BP if the number of layers becomes too large.

\subsection{Some problem-inspired ansatzes}

Just like hardware efficient is used for a wide-range of problem-agnostic circuits, problem-inspired ansatzes is a blanket term  used for models which  embed inductive biases about the problem at hand into the circuit architecture itself. In most cases, problem-inspired ansatzes adjust circuit expressiveness by using specific sets of gates, distinct entangling gate topologies, or by aligning parameters with the symmetries of the task. Here, one can generally expect that $\mathfrak{g}$ will be smaller than  $\mathfrak{su}(2^n)$~\cite{ragone2023unified}. While we will present below circuits which can in fact sufficiently reduce the circuit expressive power to potentially avoid BPs, we here discuss problem-inspired architectures that can still have exponentially concentrated loss functions despite their inductive biases. 

\textit{Hamiltonian Variational Ansatz (HVA)}. HVAs~\cite{wecker2015progress,wiersema2020exploring,larocca2021diagnosing,park2023hamiltonian} were originally introduced within the context of the variational quantum eigensolver~\cite{peruzzo2014variational}. Here,  the goal is to find the ground state of a Hamiltonian $H$ which we assume can be expressed as a sum of non commuting terms $H=\sum_s H_s$ (i.e., $[H_s,H_{s'}]\neq 0$ if $s\neq s'$). The HVA circuit is expressed as $U(\thv)=\prod_{l}\prod_s e^{-i \theta_{ls}H_s}$, from where we can clearly see that the ensuing DLA will strongly depend on what the Hamiltonian of interest $H$ is. As such, the presence or absence of BPs will be extremely problem-dependent and a case-by-case analysis will be needed. For instance, recent results have shown that the vast majority of one-dimensional translational invariant Hamiltonians $H$ will lead to exponentially large algebras and thus to BPs~\cite{wiersema2023classification}.

\textit{Quantum Alternating Operator Ansatz (QAOA).} The QAOA~\cite{farhi2014quantum,hadfield2019quantum} is widely used to variationally attempt to solve combinatorial optimization problems. Here, one  wishes to find the ground state of a Hamiltonian $H_P$ that is diagonal in the computational basis and which encodes in its ground state the solution to some combinatorial problem of interest. The QAOA is constructed as $U(\thv)=\prod_{l}e^{-i \theta_{lM}H_M}e^{-i \theta_{lP}H_P}$, where $H_M$ is the so-called mixer Hamiltonian~\cite{bartschi2020grover,fuchs2022constrained,tate2021classically} usually taken to be $H=\sum_{j=1}^n X_j$. The presence of BPs in QAOA for maximum-cut  tasks  has been analyzed in~\cite{larocca2021diagnosing,zhang2021quantum,kazi2022landscape}, where it has been shown that the loss will exhibit a BP for the vast majority of graphs if the circuit is deep enough.
Nevertheless, it is an open question the  depth for which QAOA can obtain useful solution and outperform state-of-the-art classical methods~\cite{crooks2018performance,akshay2020reachability, boulebnane2024solving,montanaro2024quantum}.
Moreover, this exponential concentration phenomenon has been also studied in QAOA variants such as the multi-angle QAOA where every gate in the circuit is assigned its own parameter~\cite{herrman2022multi}. Here, one can show that for all graphs with non-trivial cut value the circuit's DLA dimension will be exponentially large implying that the loss will be concentrated~\cite{kazi2022landscape,aguilar2024full,chapman2020characterization,kokcu2024classification}. Notably, while the depth at which BPs will occur depends on the graph, and will likely require a polynomial number~\cite{ragone2023unified,haah2024efficient}, there exists cases where QAOA can have BPs even with a depth of one~\cite{kazi2022landscape}. With the previous being said, there have been promising research direction to train QAOA in regimes where the BP proofs do not hold (e.g., training classically on small problems and transferring parameters) ~\cite{zhou2021quantum,chai2022shortcuts, chandarana2022digitized, vizzuso2024convergence,shaydulin2023parameter,farhi2022quantum}. However, the performance of these approaches at scale is still an open question.

\textit{Unitary Coupled Singles and Doubles (UCCSD) ansatz}. The UCCSD is a problem inspired PQC used to find the ground state of a fermionic molecular Hamiltonian via the variational quantum eigensolver~\cite{taube2006new,peruzzo2014variational,lee2018generalized}. Here, it has been shown that standard versions of UCC composed of controlled single excitation gates are particle-conserving
 universal unitaries~\cite{arrazola2022universal}. Indeed, the results in Ref.~\cite{arrazola2022universal} implies that deep randomly initialized UCCSD ansatzes will exhibit a BP if the state of interest lies within a subspace with a number of excitations that scales with the problem size (see~\cite{mao2023barren} for a numerical verification of this fact). In the next section we discuss how some of these issues could be avoided, e.g., by using shallow circuits, and good initializations strategies.

\section{Strategies to avoid or mitigate BPs}\label{sec:do-work}

Since the discovery of BPs, attempts have been made to mitigate their detrimental effects. A common feature shared between the tools discussed here is that they break the assumptions which lead to the curse of dimensionality. First, we will review techniques which somehow transform the loss into the inner product between two objects that live in polynomially, rather than in exponentially, large spaces~\cite{cerezo2023does} (see Box~\ref{box:subspaces}). While these approaches allow us to avoid BPs, we note that they also make variational models vulnerable to classical simulation (see more discussion in Sec.~\ref{sec:beyond}). Second, we will review methods that mitigate BPs even when operating in exponentially large spaces.  Here, we particularly refer the reader to the alternative initialization section, as this appear to be one of the most promising ways to jump start the training in a region of large gradients, rather than in a random place of the optimization landscape. 

\begin{mybox}[label={box:subspaces}]{Avoiding BPs by working in small spaces} 
Recently it has been noted that several strategies to avoid BPs do so by effectively encoding the relevant dynamics in the loss function to a polynomially-large subspace of the operator space, thus avoiding the curse of dimensionality that causes BPs. As discussed in the main text,  we can think about the adjoint action of a unitary PQC over the initial state $U(\thv)\rho U\ad(\thv)$, or the measurement operator $U\ad(\thv)O U(\thv)$ as only being able to exactly, or mostly, explore a polynomially large subspace $\BC_\lambda$ of $\BC(\HC_0)$. That is $\dim(\BC_\Lambda)\in\OC(\poly(n))$ Denoting  $A^{\lambda}$ as the projection of an operator onto  $\BC_\lambda$, the loss function becomes $\ell_{\thv}(\rho,O)=\langle \rho_{\thv}^\lambda,O^\lambda\rangle=\langle \rho^\lambda,O_{\thv}^\lambda\rangle$. Here, it is extremely important to note that $\BC_\lambda$ is not necessarily a module of the Lie group associated with the quantum circuit. As such, the loss function variance will not always  be given by Eq.~\eqref{eq:var-DLA}. Still, since we can only move in a small subspace, we can expect that when one varies the parameters the variance can decay polynomially as $\Omega(1/\dim(\BC_\lambda))$. Of course, the input state and the measurement operators still need to be well aligned with $\BC_\lambda$, as if either $\rho$ or  $O$ have exponentially small component in the subspace, the loss will be deterministically concentrated. 
\end{mybox}

\subsection{Shallow circuits}

Perhaps one of the simplest ways to avoid BPs is to employ shallow PQCs.  Here, even if the loss is  concentrated for sufficiently deep versions of the circuit (either because it is too expressive or due to the effects of noise), if the number of gates is sufficiently small one can expect that that the effect of both probabilistic and deterministic BPs is  reduced for certain measurements. In addition, while this strategy does not help circumventing the deterministic BP arisen from the input state, the shallow depth PQCs can avoid the (unital) noise-induced BP. Examples include logarithmic-depth hardware-efficient ansatzes with local measurements~\cite{cerezo2020cost,leone2022practical,zhang2023absence},  or quantum convolutional neural networks~\cite{pesah2020absence,zhao2021analyzing} which are shallow by design (but also classically simulable~\cite{bermejo2024quantum}). Intuitively,  these circuits avoid BPs---even if randomly initialized---by constraining the space that the measurement operator can explore. These insights have been used in the literature to show that when using the variational quantum eigensolver, along with a shallow circuit, using problem-encoding schemes that lead to more local measurement operators is more effective~\cite{uvarov2020barren,cichy2022perturbative}.

It is important to note that while shallow circuits can provably have non-exponentially vanishing gradients, this is only true for certain measurements. For instance, a logarithmic-depth hardware efficient ansatz can still exhibit a BP for a non-local measurement~\cite{cerezo2020cost}. The question still remains as to whether a good  solution exists within the set of parameters. By using few gates, it is entirely possible that the set of parameters obtained from the optimization does not correspond to a good enough solution of the problem~\cite{larocca2021theory}. This phenomenon is known as a \textit{reachability deficit}~\cite{crooks2018performance,akshay2020reachability}. In general, there is no simple way of knowing whether the model will suffer from reachability deficits and a case-by-case analysis is necessary. In addition, it has also been reported that shallow circuits can have spurious local minima~\cite{kiani2020learning,larocca2021theory,rajakumar2024trainability} making it even harder to find any good solution that might, or not, exist in the landscape.

\subsection{Small dynamical Lie algebras}

In the previous section, we presented problem-inspired ansatzes that are still too expressive to avoid exponential concentration. However, there are cases of PQCs whose DLA dimensions grow only polynomially with the number of qubits. In these cases, depending critically on the choice of initial state and measurement operator, the circuit’s adjoint action may result in small modules. This allows the circuit to avoid probabilistic expressiveness-induced BPs, even in its deeper forms, as described by Eq.~\eqref{eq:var-DLA}. 

The circuit’s expressiveness can be reduced by using circuits with fewer qubits~\cite{cichy2022perturbative,marshall2022high,zhang2020toward}, or encoding sufficient inductive biases about the problem. In this context, it has been recently pointed out that one particularly important source of inductive biases is the set of symmetries that a problem satisfies. That is, if  one knows that the task's solution must respect some symmetry, then one should construct a variational models that only explore the symmetry-respecting solution space~\cite{larocca2022group,ragone2022representation,nguyen2022atheory,meyer2022exploiting,skolik2022equivariant,cerezo2022challenges,volkoff2021large,sauvage2022building}.  For instance, it has been shown that permutation-equivariant unitary PQCs --circuits that remain invariant under qubit permutations-- have a DLA whose dimension scales as $\OC(n^3)$~\cite{schatzki2022theoretical,kazi2023universality}. Similarly,  $U(1)$-equivariant circuits, have been shown to avoid BPs when the Hamming weight of the initial state does not scale with $n$~\cite{larocca2021diagnosing,monbroussou2023trainability,cherrat2023quantum}. To finish, we note that~\cite{lee2021towards} studied a PQCs where all the gates commute, meaning that the size of the algebra is exactly equal to the number of distinct gate generators.

\begin{mybox}[label={box:HEA-shallow}]{Its all relative (to  the modules)}
Until very recently, the BP study was performed on a case-by-case basis, analyzing one architecture at a time. While such fragmented patchwork analysis was pivotal to the recent development of a  unified theory of BPs~\cite{ragone2023unified}, it lead to several misconceptions in the literature where lessons learnt in one architecture are taken to be generically true and extrapolated to other PQCs. These include claim such as ``global'' measurements such as $O_G=X_1Z_2\cdots X_{n-1}Z_{n}$ (i.e., when $O$ act nontrivially on all the qubits) lead to BPs, while local measurements such as $O_L=X_{n/2}$ (i.e., we measure just one qubit) prevent them. While these claims are indeed true for certain architectures such as logarithmic depth hardware efficient ansatzes~\cite{cerezo2020cost,marrero2020entanglement,leone2022practical} (or to the single-qubit rotation ansatz of Box~\ref{box:single-qubit}), they are certainly not generically true. For instance, a parametrized matchgate circuit can avoid BPs with $O_G$, but will lead to exponentially concentrate loss functions if we pick $O_L$~\cite{diaz2023showcasing}. To understand why those two cases behave so differently, we need to (again) study the module, or subspace, that the adjoint action of the PQC leads to when action over $O$ (see also Box~\ref{box:subspaces}). For the single qubit rotation ansatz, $O_G$ belongs to an exponentially large module, while $O_L$ to a polynomially small one. This statement is fully reversed in the matchgate case as $O_G$ is in a small module, but   $O_L$ in an exponentially large one.    
\end{mybox}

Another archetypal example of a PQC with small DLA is that of  parametrized matchgate circuits (also known as Ising-model HVAs~\cite{jozsa2008matchgates,wan2022matchgate,de2013power,oszmaniec2022fermion,cherrat2023quantum, matos2022characterization, diaz2023parallel}), where one can find that $\mathfrak{g}\simeq\mathfrak{so}(2n)$ and therefore $\dim(\mathfrak{g})\in\OC(n^2)$~\cite{kokcu2022fixed,wiersema2023classification,larocca2021diagnosing,diaz2023showcasing} (see Box~\ref{box:HEA-shallow}). These circuits can be shown to have a reduced expressive power as they can only implement  free-fermionic evolutions. Along a similar line, it has been shown  that linear optics continuous variable PQCs acting on coherent light in $n$ modes have an associated algebra $\mathfrak{g}\simeq\mathfrak{o}(2n)$~\cite{volkoff2021efficient,Volkoff2021Universal} thus also potentially leading to BP-free loss functions. Interestingly, these models have also been explored in classical machine learning, where they are known as efficient unitary neural networks~\cite{arjovsky2016unitary,jing2017tunable}, and where it was already noted that they do not exhibit vanishing gradients at large depths.  We note, however, despite being able to avoid BPs, working with the polynomial DLA dimension could lead to the classical simulability of the VQAs (see Sec.~\ref{sec:beyond}).

While the previous examples illustrate that one can indeed find circuits with polynomial-sized DLAs, and thus to their action being reducible to some polynomial-sized space, it is worth noting that such cases are quite rare. In fact, as previously mentioned most circuits will exhibit exponentially large algebras~\cite{wiersema2023classification,larocca2021diagnosing,kazi2022landscape}, and the quest for other PQC architectures with small algebras is an open area of research.

\subsection{Variable structure ansatzes}

Variable structured PQCs attempt to mitigate probabilistic expressiveness and noise induced deterministic BPs by leveraging classical machine learning protocols to iteratively grow the  quantum circuit by placing or removing gates that lower the loss function and keep large gradients. As such, these  approaches are able to employ highly-expressive gate-sets by only exploring specific regions
of the ansatz architecture hyperspace. Several machine learning-aided strategies have been applied, including evolutionary algorithms, short-depth compilation algorithms, and automachine learning~\cite{bilkis2021semi,du2020quantum,grimsley2019adaptive,tang2019qubit,sim2021adaptive,zhang2021mutual,tkachenko2020correlation,claudino2020benchmarking,rattew2019domain,chivilikhin2020mog,cincio2018learning,zhang2020differentiable,wada2022sequential}. In particular, we highlight that one of the most-promising and well-studied variable ansatz strategy is the ADAPT-VQE (and variants there-off for other circuits such as QAOA)~\cite{tang2019qubit,grimsley2019adaptive,grimsley2022adapt,anastasiou2024tetris,van2024scaling,romero2022solving,zhu2022adaptive}. For instance, one can  construct a circuit for quantum chemistry applications with gates taken from some poot (e.g., from those arising in  UCCSD), and it has been shown that this strategy has led to hardware-friendly circuits with few entangling gates that focus on regions with large gradients, often resulting in good solutions.

\subsection{Alternative initialization strategies}\label{sec:initialization-strategies}

One of the main limitations of studying probabilistic BPs by analyzing the scaling of $\Var_{\thv}[\ell_{\thv}(\rho,O)]$ is that these quantities only tell us how concentrated the loss is on average across the \textit{whole} landscape.  Such information is of course important as it allows us to understand what to expect if we randomly initialize the model's parameters. However, it is well known that randomly initializing the parameters is never a good idea, and that the initialization method can be critical in determining the model’s ultimate performance. 

Motivated by this observation, several strategies to initialize a variational quantum computing scheme have been proposed.  These include restricted  small angle initializations (where one still randomly initializes the parameters but in a more structured way, such as in a special region)~\cite{grant2019initialization,jain2022graph,park2024hardware,skolik2020layerwise,zhang2022escaping,kulshrestha2022beinit,rad2022surviving,astrakhantsev2023phenomenological,wang2023trainability,haug2021optimal,kashif2023alleviating,shi2024avoiding}, pre-training via classical~\cite{friedrich2022avoiding,grimsley2019adaptive,rudolph2022synergy,marin2021quantum} or quantum~\cite{cervera2020meta,goh2023lie} methods, parameter transfer~\cite{brandao2018fixed,zhou2020quantum,wurtz2021fixed,boulebnane2021predicting,galda2021transferability,farhi2022quantum,shaydulin2023parameter,mele2022avoiding,liu2023mitigating} and iterative learning strategies~\cite{valls2024variational}. For the latter case, we note that analogous sample complexity lower bounds have been derived for alternative iterative variational approaches via the McLachlan principle~\cite{Zhang2023LowDepth,Gacon2024variational}.

Here we highlight that some of these initialization strategies have shown heuristic success and appear to be one of the most promising avenues to mitigate BPs. In fact, classical machine learning models can exhibit vanishing gradients issues just like their quantum counterpart (see below for more on this), and smart-initializations have been one of the main tools used therein. Additionally, in chemistry problems such as finding a ground state of a molecular Hamiltonian, the initialization is done typically by taking into a problem structure such as initializing with a mean field approach. Hence, we expect that warm-start will play a fundamental role in the future of variational quantum computing. However, these strategies have limitations that must be considered~\cite{campos2021training,campos2021abrupt}. These include concerns such as the fact that one can initialize in a region which might have large gradients, but is not well connected to a global minima, or that might be classically simulable~\cite{cerezo2023does}.  Indeed several workarounds have been proposed to prevent a smartly initialized circuit to run into problematic regions of the optimization landscape (such as monitoring the reduced state's entanglement~\cite{sack2022avoiding}, or using state-of-the-art  optimizers~\cite{fitzek2023optimizing,acampora2023comparison,bermejo2024improving}).

\section{Strategies that cannot avoid BPs}\label{sec:not-work}

While many promising strategies have been devised to prevent or alleviate BPs, some intuitive approaches fail because they do not address the underlying causes of BPs. Here we review two such strategies.

\textit{Changing the optimization method.}
It is common to find claims that it is possible to navigate a BP landscape (when randomly initialized) by changing the optimization method. For example, one might assume that despite suppressed gradients, it is possible to find a loss-minimizing direction by using higher-order gradient information, gradient-free methods, quantum natural-gradients~\cite{stokes2020quantum,koczor2019quantum} or other quantum fisher information-based schemes.  However, studies have shown that these approaches still require an exponential number of measurements to obtain a loss-minimizing direction~\cite{cerezo2020impact,arrasmith2020effect,thanasilp2021subtleties}.

\textit{Error mitigation and noise-induced BPs.} Given the critical limitations that noise imposes on variational quantum algorithms, it is essential to investigate whether error mitigation techniques~\cite{temme2017error, li2017efficient, cai2022quantum} can sufficiently denoise the loss function to find a loss-minimizing direction. Unfortunately, research has shown that in the worst cases, mitigating noise-induced BPs requires an exponential number of resources~\cite{franca2020limitations,wang2021mitigating,takagi2021fundamental,takagi2022universal}. This remains true even when the circuit has sub-logarithmic depth in certain worst-case scenarios~\cite{quek2022exponentially}. Importantly, we note that it is not known whether such worst-case scenario is representative of average-cases.

\section{Exponential concentration elsewhere}\label{sec:impact}

The study of BPs has both informed and been influenced by other areas of quantum computing and machine learning. Here we briefly review some of these results.

\textit{Quantum generative models.} 
In quantum generative modeling one aims to optimize a quantum state such that its probability distribution (resulting from some set of measurements) matches a target distribution~\cite{perdomo2018opportunities, gao2022enhancing, benedetti2019generative, liu2018differentiable, coyle2020born, rudolph2023trainability, zoufal2021generative, lloyd2018quantumgenerative, letcher2023tight, Amin2018Quantum, coopmans2023sample, chang2024latent}.
Quantum generative models are implicit because they only provide access to samples, unlike explicit models that provide direct access to the underlying distributions. Correspondingly, one can draw a distinction between explicit losses, which are formulated explicitly in terms of the model and target probabilities, and implicit losses, which compare samples from the model and the training distribution~\cite{rudolph2023trainability}. The tension between using an explicit loss, such as the Kullback-Leibler divergence, with an implicit model, was shown in~\cite{rudolph2023trainability,leadbeater2021f} for quantum circuit Born machines to lead to a BP. 
Conversely, implicit losses, such as the maximum mean discrepancy~\cite{liu2018differentiable, rudolph2023trainability} and some losses used for quantum generative adversarial networks~\cite{letcher2023tight,leadbeater2021f,chang2024latent}, can be viewed as the expectation value of an observable and so can be analyzed with the tools discussed previously.

BPs have also been studied in quantum Boltzmann machines~\cite{Amin2018Quantum},  where a parametrized thermal quantum state model is trained by minimizing a log-likelihood loss. In this case, the landscape is convex and so it is possible to derive convergence guarantees with only polynomial sample complexity~\cite{coopmans2023sample}. Inherently quantum strategies for generative modelling also pose an interesting avenue for finding quantum generative models that avoid BPs~\cite{kieferova2021quantum, rudolph2023trainability}. 

\textit{Kernel-based quantum algorithms.}  Exponential concentration also arises in kernel-based schemes~\cite{huang2021power, thanasilp2022exponential, kubler2021inductive, shaydulin2021importance, canatar2022bandwidth,  suzuki2023effect, suzuki2022quantumfisher, xiong2023fundamental,yu2023expressibility}, where a quantum computer is used as an enhanced feature-space from which inner products or similarity measures between the data states can be estimated (\textit{a la} kernel trick)~\cite{schuld2018supervised, havlivcek2019supervised,schuld2021quantum,gan2023unified}. Unlike PQCs, one strength of kernel methods is that for a given choice of a kernel function and dataset, one is guaranteed to find the optimal trained model that minimizes the empirical loss since the training landscape is convex.  
However, the kernel values can only ever be estimated statistically using a polynomial number of shots. If the kernel is exponentially concentrated, then the empirically estimated kernels contain, with high probability, no meaningful information about the input data. Training the model with these estimated kernel values then leads to a trivial data-insensitive model where the predictions are independent of the input data~\cite{thanasilp2022exponential}.

\textit{Quantum optimal control.} In quantum optimal control, exponentially concentrated optimization landscapes can arise~\cite{kiani2020learning}, but their connection to BPs was only explored after the BP literature gained traction. The intrinsic connection between these two areas of study was first reported in~\cite{magann2021pulses} and later in~\cite{ge2022optimization}. Then, in~\cite{larocca2021diagnosing} it was  noted that  tools from quantum optimal control, such as the study of the PQC's DLA, can be used to diagnose BPs. 
Nowadays, some tools developed in variational quantum computing have started to see use in optimal control schemes~\cite{broers2024mitigated,tao2024unleashing,dekeijzer2023pulse,ge2022optimization,
broers2024mitigated,
pecci2024beyond}, and the extent of this cross-fertilization is yet to be seen. 

\textit{Trainability of tensor networks.} Several works have studied BPs in tensor networks. For instance, it has been shown that tensor network based machine learning models exhibit BPs for global but not local losses~\cite{liu2021presence}, and isometric tensor network states can avoid BPs~\cite{miao2023isometric}. Similarly, BPs have been studied in  quantum circuits inspired by matrix product states, tree tensor networks, and the multiscale entanglement renormalization ansatz~\cite{martin2022barren,barthel2023absence}.

\textit{Learnability.} 
Insights from BPs have allowed researchers to present theorems on the difficulty of variationally learning typical random states~\cite{cerezo2020cost}, typical random unitaries~\cite{holmes2020barren,garcia2023barren}, and the output distributions of random circuits~\cite{rudolph2023trainability}. 
There is an interesting commonality between these results and statistical query learning~\cite{Feldman2016Statistical, arunachalam2020quantum} no-go theorems for learning typical random states~\cite{anshu2024survey}, random unitaries~\cite{angrisani2023learning, wadhwa2023learning, nietner2023unifying}, the expectation of an observable over random states~\cite{anschuetz2022beyond} and the outputs distributions of random quantum circuits~\cite{nietner2023average}. In all cases, it is worth noting that while BPs are fundamentally an algorithm-dependent phenomenon (but are, in some sense, independent of the learning problem), tools from statistical query learning can provide algorithm-independent bounds on the hardness of learning a specific function class. Nonetheless, mathematically there are close links between the two families of results, with the difficulty in both cases boiling down to a curse of dimensionality. For a thorough investigation of the link between barren plateaus and statistical query learning no-go theorems see Ref.~\cite{nietner2023unifying}.

\section{Beyond barren plateaus in variational quantum computing}\label{sec:beyond}

The study of BPs does not fully address the challenges in ensuring that variational quantum computing models can achieve a quantum advantage.  Here we briefly review some of the other questions that need to be addressed.

\textit{Local minima \& solution quality.} It has been shown that quantum landscapes can be plagued with exponentially many sub-optimal local minima, potentially making their optimization extremely hard~\cite{anschuetz2022quantum,anschuetz2021critical,bittel2021training,you2021exponentially,fontana2022nontrivial,rajakumar2024trainability,huang2024learning,nemkov2024barren,bermejo2024improving, anschuetz2024unified,nemkov2024barren}. Critically, the local minima issue affects both shallow circuits, where no good solutions exist, and deep circuits, where a solution might exist but remains computationally difficult to reach. For example, when learning an unknown unitary with a shallow one-dimensional parametrized circuit, there are exponentially many sub-optimal local minimum~\cite{huang2024learning}. Indeed, we know that if the PQC is too shallow, then spurious local minima can appear in the landscape as an effect of not having enough parameters to control all available directions. While these  minima can be  be removed  by overparametrizing the PQC, i.e., by adding more parameters, this could requires exponentially deep circuits~\cite{kiani2020learning,larocca2021theory, campos2021abrupt,tikku2022circuit, anschuetz2024unified}. Such realization actually leads to a very powerful insight: If we have a randomly initialized circuit with $\OC(\poly(n))$ parameters which has a BP because the DLA is exponentially large (and thus the PQC is not overparametrized~\cite{larocca2021theory}), even if we allow for infinite measurements and get rid of the sampling noise, we will still not be able to train since the optimizer is exponentially likely to get stuck in a local minima~\cite{kiani2020learning,larocca2021theory, campos2021abrupt,tikku2022circuit,nemkov2024barren}. Addressing the problem of local minima will be fundamental to guarantee that a model will be useful (this will be particularly important when using alternative-initialization techniques).

\textit{Classical simulability of BP-free models.} As we have discussed above, many existing techniques that avoid the BP curse of dimensionality do so by encoding the relevant dynamics in some subspace that grows polynomially in the number of qubits $n$. Since an exponentially large space is no longer being used, one should be able to simulate the information processing ability of the PQC, and thus classically estimate the loss function~\cite{cerezo2023does} (note that for non-classically simulable initial states or measurements the quantum computer may be needed to gather some data in an initial data collection phase). This is possible for all parameter settings for shallow hardware efficient ans\"{a}tze~\cite{basheer2022alternating, jerbi2023power} and problem inspired ans\"atze with polynomial DLAs~\cite{somma2005quantum, somma2006efficient, galitski2011quantum, goh2023lie,anschuetz2022efficient}, and possible for any randomly chosen parameter setting with high probability for a wide class of unstructured parameterized quantum circuits~\cite{angrisani2024classically, rudolph2023classical}, QCNNs~\cite{bermejo2024quantum}, noisy circuits~\cite{fontana2023classical, mele2024noise} and small-angle initialization strategies~\cite{lerch2024efficient}.

Not having to implement a PQC on the quantum device could lead to algorithms that are better suited for near-term quantum computers as the data acquisition phase could be less noisy than fully implementing the parametrized quantum circuit~\cite{cerezo2023does,bharti2020iterative,bharti2020quantum}. Moreover, the connection between classical simulability ans trainability is a current area of analysis~\cite{gil2024relation}.  Elsewhere, recent results have showcased that classical algorithms, when equipped with measurements from a quantum computer, can be significantly more powerful than their fully classical counterparts when solving certain contrived tasks~\cite{parrish2019quantum, huang2021power, jerbi2023shadows, gyurik2023limitations,elben2022randomized}. Still, while it remains to be  determined if such separation holds for practical applications, this is regardless a promising direction of research.

Finally, it is worth highlighting that there are numerous fully classical machine learning approaches that tackle the same applications as variational quantum computing. These include variational tensor network methods~\cite{white1992density, lee2023evaluating} and neural network states~\cite{carleo2017solving, sharir2020deep, carrasquilla2017machine, torlai2018neural} for solving ground state problems, simulating the dynamics of quantum systems and state tomography. While limited by the intrinsic challenges of classically simulating quantum systems, these approaches naturally sidestep the barren plateau problem completely.

\section{Link with vanishing gradient problem}\label{sec:classical}
A common question when studying BP phenomena is the relationship to vanishing gradient problems in classical deep learning.  Identifying a precise connection between these vanishing gradients has been challenging. This is partly because there are several different types of vanishing gradients in classical settings. Here we review some of the potential similarities and differences that may lead to a multitude of perspectives. 

\textit{Differences between the two phenomena.}
The most well-known variety of classical vanishing gradients is the vanishing or exploding norm with \textit{depth}, caused by growing or shrinking weights due to the activation functions~\cite{glorot2010understanding}. This phenomenon differs from BPs in variational quantum computing, where (non-noise-induced) BPs occur as the number of qubits increases, not with circuit depth. That is, while quantumly we have a `curse of dimension', classically the problem would more accurately be described as a `curse of \textit{depth}'. In addition, it is worth noting that in the quantum setting, the BP phenomenon can be intimately tied to what $\rho$ and $O$ are, as different initial states and measurements can lead to a model having (or not having) exponentially concentrated loss functions~\cite{cerezo2020cost,diaz2023showcasing}.

\textit{Classical solutions to vanishing gradients.} 
Many solutions to the vanishing of classical gradients with depth have been developed, including batch normalization~\cite{ioffe2015batch}, layer-wise normalization~\cite{ba2016layer} and the use of Re-LU activation functions~\cite{nair2010rectified}, although it is worth mentioning that these techniques might still fail in some situations~\cite{hochreiter1998vanishing,hochreiter1997long}. As these solutions target the problem of shrinking weights at large depths due to the non-linearity of the activation function, these solutions seem not be directly applicable to the quantum case.  

Some deep classical neural networks still suffer problems even with the solutions listed above. In models like recurrent neural networks, these effects can be examined as the input length increases to infinity, and one can analyze contributions to the current prediction from the infinite past. Unrolling this network leads to a model exhibiting infinite depth, where the problems of vanishing gradients are only further magnified~\cite{hochreiter1998vanishing}. For a signal to survive from the infinite past, one has to conceive of unitary neural networks that conserve signal at every step, similar to the unitary operations of quantum computers~\cite{jing2017tunable}.  The general strategy of preserving information from earlier / past layers to improve gradient signals is applicable to networks beyond RNNs, and these solutions include skip-connections, residual connections, and general LSTM models~\cite{hochreiter1997long}. Strategies derived from the idea of carrying information forward more robustly could be emulated using simultaneous preparations of quantum states, but direct copying is challenging due to the no-cloning theorem.

Finally, restricted angle range initialization strategies~\cite{Narkhede2022Review, Chen2018Dynamical, Xiao2019Dynamical} and pre-training methods~\cite{saxe2013exact} have proven effective classically for avoiding the vanishing gradient problem. Analogous methods have been shown to lead to larger gradients in quantum contexts~\cite{zhang2022escaping,park2023hamiltonian, wang2023trainability, park2024hardware, shi2024avoiding, rudolph2022synergy}. However, given that quantum landscapes have been shown to have numerous local minima in under-parameterized regimes~\cite{anschuetz2022beyond} this strategy might be seen as less promising in quantum contexts.

\textit{The role of exponentially large spaces.} A common refrain when discussing concentration of measure is that the dimension of quantum spaces can be so much larger than classical spaces. However, there are subtleties here, as classical models can also explore exponentially large spaces  (e.g., one can classically optimize a $2^n$-dimensional probability distribution on $n$ bits via probabilistic parametrized bit-flips). Thus, it seems that it is not the size of the space explored by quantum models \textit{per se} but rather \textit{how} the space is used. Indeed, it is the hope of quantum computing that we may find novel, less computationally intense, paths through similar spaces of problems. Hence, it appears that a key step forward in resolving the problem of BPs is not just a powerful ansatz, but a parametrization that has a natural bias towards problems of interest.

\textit{The cost of precision.}
In the classical case, the value of the loss and gradient can often be obtained to precision $\epsilon$ with resources that scale as $\OC(\log(1/\epsilon))$, so that even if a gradient is rapidly vanishing, we may compensate  with additional precision.  In contrast, for most quantum models based on expected values, we are relegated to stochastic sampling at a cost of $1/\epsilon^2$ or $1/\epsilon$ with clever use of amplitude amplification. This means that any problems of decaying gradients are often exponentially magnified in the quantum case. However, there are classical models in use that make use of stochastic sampling to determine the value of a loss and gradient~\cite{baird1998gradient,salimans2017evolution}, so quantum models are not entirely alone in this difficulty. An interesting strategy to sidestep this problem in quantum models could be to take inspiration from the classical side and move away from expected values to most-likely readout values or another deterministic output that can be reliably determined to high precision without sampling.

\section{Implications and outlook}\label{sec:outlook}

The community's efforts have led to significant advancements in understanding the BP phenomenon.  The primary causes of the phenomenon, which in essence all stem from a curse of dimensionality, have been pinned down. Similarly, progress has been made in identifying architectures that circumvent BPs. However, the ultimate impact of the BP phenomenon on the scalability  of variational quantum models is still to be determined.

The questions that remain largely boil down to the fact that the BP phenomenon is an average case notion. If a landscape has a BP then the probability that any randomly chosen parameter setting has non-negligible loss gradients is exponentially small. Given that quantum loss functions are precision limited, i.e., they are estimated from a finite number of measurement shots, the probability of obtaining an informative signal at any randomly chosen parameter setting is exponentially small. Nonetheless, in exponentially narrow regions, a BP landscape can exhibit substantial gradients. The question is whether these regions are easy to find, and even if we find them whether they are useful and can allow for meaningful training. Addressing this requires understanding the interplay between BPs, local minima, expressivity limitations, and, crucially, classical simulability; all of which are open questions that we urge our community to explore.  Indeed, we believe that the field of variational quantum computing could benefit from taking a holistic view of all these difficulties and the underlying causes that underpin them. 

At this point, we are thus faced with the question of \textit{what next}? Classical machine learning has been driven by a combination of rigorous theory and well-informed heuristic strategies, leading to successes beyond those predictable by theory alone. 
However, we do not yet have the quantum hardware needed to perform large-scale heuristic implementations of variational quantum models. Indeed, the original motivation for studying BPs was to understand limitations that could prevent scaling  variational models to non-classically simulable regimes. In this regard, the field of BP  has  successfully carved out settings where we know a quantum advantage will be unlikely. Still, much work remains to be done. We  envision that the BP study will continue to serve as a guide to determine if, and where, a practical variational quantum advantage exists, and to hint towards the development of new variational methods that go beyond the standard procedure of Fig.~\ref{fig:fig-1}.

\medskip

\section*{Acknowledgements}

We thank Maria Kieferova, Pablo Bermejo, and 
Tom O'Leary for their feedback on our manuscript. 
 M.L. was supported by the Center for Nonlinear Studies at Los Alamos National Laboratory (LANL). M.L. acknowledges support by the Laboratory Directed Research and Development (LDRD) program of LANL under project number 20230049DR. 
S.T. and Z.H. acknowledge support from the Sandoz Family Foundation-Monique de Meuron program for Academic Promotion. 
This research was partly supported (L.C.) by the Quantum Science Center, a National Quantum Science Initiative of the Department of Energy, managed by Oak Ridge National Laboratory. L.C. was also supported by the U.S. Department of Energy, Office of Science, Office of Advanced Scientific Computing Research, under Computational Partnerships program. M.C. acknowledges support by LANL ASC Beyond Moore’s Law project and by LDRD program of LANL under project number 20230527ECR.

\end{document}